\documentclass[a4paper,11pt]{article}
\usepackage{jheppub}
\makeatletter
\renewcommand{\@fpheader}{}
\makeatother
\usepackage{todonotes}
\usepackage{float}
\usepackage{graphicx}
\usepackage{soul}
\usepackage{comment}
\usepackage{enumitem}

\usepackage{amsmath,amssymb}  
\usepackage{xcolor}
\definecolor{darkpurple}{rgb}{0.5, 0.2, 0.8}
\definecolor{darkblue}{rgb}{0.0, 0.0, 0.8}
\definecolor{darkgreen}{rgb}{0.0, 0.4, 0.0}
\definecolor{darkred}{rgb}{0.5, 0.0, 0.0}
\usepackage{hyperref}
\hypersetup{
    linktocpage,
    colorlinks,
    citecolor=darkgreen,
    linkcolor= darkgreen,
    urlcolor=darkgreen,
    filecolor=darkgreen
}

\definecolor{darkgreen}{rgb}{0.0, 0.4, 0.0}

\newcommand{\cB}{{\mathcal B}}
\newcommand{\RR}{ {\mathbb R}}
\newcommand{\cO}{ {\mathcal O}}
\newcommand{\rI}{ \text{I} }
\newcommand{\ii}{ \text{I}\hspace{0.1em}\overline{\text{I}}}

\def\dimO{\Delta_{\mathcal{O}}}
\def\dimOi{\Delta_{\mathcal{O}_i}}

\newcommand{\im}{ {\text{Im} }}
\newcommand{\re}{ {\text{Re} }}

\title{Renormalons as Saddle Points}

\author{Arindam Bhattacharya,}
\author{Jordan Cotler,}
\author{Aurélien Dersy and }
\author{Matthew D.  Schwartz}
\affiliation{Department of Physics, Harvard University}

\emailAdd{arindamb@g.harvard.edu}
\emailAdd{jcotler@fas.harvard.edu}
\emailAdd{adersy@g.harvard.edu}
\emailAdd{schwartz@g.harvard.edu}

\abstract{
Instantons and renormalons play important roles at the interface between perturbative and non-perturbative quantum field theory. They are both associated with branch points in the Borel transform of asymptotic series, and as such can be detected in perturbation theory. However, while instantons are associated with non-perturbative saddle points of the path integral, renormalons have mostly been  understood in terms of Feynman diagrams and operator product expansions. We suggest a non-perturbative path integral explanation of how both instantons and renormalons produce singularities in the Borel plane using representative finite-dimensional integrals. In particular, we build evidence that renormalons can be understood as saddle points of the 1-loop effective action, enabled by a crucial contribution from the quantum scale anomaly. These results are illustrated in simple toy models and indicate a possible route toward studying renormalons within realistic asymptotically-free field theories.
}

\begin{document}

\maketitle
\flushbottom

\newpage
\section{Introduction}

The interface between perturbative and non-perturbative physics is at the heart of some of the deepest challenges in quantum field theory. Two objects which allow for quantitative exploration of this interface are instantons and renormalons~\cite{Lipatov:1976ny,tHooft1979,Lautrup:1977hs}. Instantons are semi-classical objects: solutions to the classical (Euclidean) equations of motion of a theory which mediate non-perturbative phenomena, such as tunneling.  Renormalons leave a similar imprint on perturbation theory as instantons but have resisted  a semi-classical interpretation. Many insights into renormalons have come from 2D models~\cite{Gross:1974jv, Novikov:1984ac,Beneke:1993yn, Marino:2019fvu, Marino:2019eym}, supersymmetric models~\cite{Schubring:2021hrw,Shifman2015,Shifman:2022xsa}, and models in compactified spacetimes~\cite{Argyres:2012vv, Argyres:2012ka, Dunne:2012zk, Dunne:2012ae}. Infrared (IR) renormalons are particularly important as they associate growth in perturbation theory with the size of power corrections~\cite{Maiani:1991az,Beneke:1998ui}. Practical uses of renormalons include motivating judicious choices of renormalization scheme~\cite{grozin1997higher,Hoang:2009yr} and heavy quark mass determination~\cite{Luke:1994xd,Beneke:2016cbu,Hoang:2017suc}.  Because instantons have a semi-classical interpretation, one can reconstruct information about the classical action and corresponding field configurations of instantons from their signature in the Borel plane.  Analogously, the Borel transform of a perturbative series containing a renormalon suggests that the renormalon should have an associated action -- but the action of \textit{what}?   We answer this question by showing that renormalons correspond to saddle points of the effective action in field theories with anomalous scale invariance.

Part of the reason renormalons are of interest is because of their importance for studying non-perturbative phenomena. When one thinks of non-perturbative physics, instantons naturally come to mind, as they are saddle points of the Euclidean action. However, in a sense renormalons are probably more important  than instantons: they correspond to stronger asymptotic growth than instantons and their characteristic scale,  for example, in QCD is  $\mu \exp(-\frac{1}{2\beta_0 \alpha_s(\mu)}) = \Lambda_{\text{QCD}}$ and exactly corresponds to where perturbative methods breaks down (in the 1-loop approximation). While instantons arise naturally from the path integral, renormalons  are typically only approached diagrammatically. In~\cite{Babansky:2000}, the authors attempted to bridge the gap between instantons and renormalons by using constrained instanton methods to show that field configurations along a valley between the vacuum saddle and instanton configurations encode the renormalon singularity in the Borel plane.  Although this approach is able to find the renormalon, it misleadingly associates the renormalon with instanton configurations and fails to identify the renormalon as a collection of saddle points. In fact, the existence of a valley and instantons are irrelevant. Our more general perspective allows for the identification of the renormalon saddles entirely independent of whether there are instantons in the theory or not.

Our approach is to begin with the more general question of how Borel transforms of asymptotic series encode the `actions' of exponential integrals.  For series associated with instantons, we show that in many cases one can fully rebuild the relevant part of the action from the series, which amounts to uncovering its non-perturbative definition. The semi-classical interpretation of renormalons is more subtle and requires consideration of Lefschetz thimble 2-manifolds embedded in $\mathbb{C}^2$. By repackaging insights from Morse theory in more physical terms, we explain an effective way to think about the multidimensional case.  We show that by judiciously integrating out degrees of freedom, operator expectation values containing renormalons can be understood with 2D integrals, and generate a renormalon pole at the expected location in the Borel plane.   More generally, we argue that for a classically scale-invariant field theory, the activation of the scale anomaly generates an effective action with renormalon saddles which are not present in the bare action. We emphasize that our work here is to identify renormalons in high order perturbative expansions in QFTs directly in the path integral as a saddle. We leave the question of precision checks of our proposal, which involves a delicate treatment of the domain of path integration, to future work.

The rest of the paper is organized as follows: In section \ref{sec:review}, we review our conventions for Borel transformations and introduce the \emph{Borel-Action} correspondence, highlighting a duality between the action of an exponential integral and its associated Borel transform. We discuss the type of singularities that one can encounter in the Borel transform and point out which ones correspond to renormalons. Following that in section \ref{sec:ren_saddle}, we elevate the general treatment of the renormalon as a saddle point in a general scale invariant quantum field theory and discuss its associated thimble. Finally, we conclude in section \ref{sec:conclusions}.

\section{Borel transforms and exponential integrals}\label{sec:review}
We begin by reviewing the concept of the Borel transform associated with a given asymptotic series. When the  series arises from an exponential integral, we highlight a duality between the corresponding action and the Borel transform -- an association we refer to as the action-Borel correspondence.  
\subsection{Action-Borel correspondence}

The Borel transform is defined from a series $f(g) = \sum_{n=0}^\infty c_n g^{n+a}$ by
\begin{equation}
  B (t) =\cB[f](t)= \sum_{n = 0}^{\infty} \frac{c_n}{\Gamma (n + a +
  1)} \,t^{n + a} \,, \label{fB}
\end{equation}
where $a$ can be any number greater than $-1$.  We will use the notations $B(t)$ and $\cB[f](t)$ interchangeably.  The corresponding inverse Borel transform is
\begin{equation}
  \cB^{- 1} [B](g) = \frac{1}{g} \int_0^{\infty} d t\  e^{- \frac{t}{g}} B
  (t)\,, \label{Binv}
\end{equation}
which formally reproduces the original series order-by-order in $n$.  We note that the function $\frac{1}{g} f (g)$ is in obvious 1-to-1
correspondence with $f (g)$; however, its Borel transform is 
\begin{equation}
 \cB \!\left[ \frac{1}{g} f (g) \right] = \frac{d}{d t} \cB [f
  (g)] \label{diffB}
\end{equation}
which is shown by differentiating Eq.~\eqref{fB}.  
This suggests that while the location of branch points in the Borel transform might have physical significance, the nature of singularities (logarithmic or power law) may not be so important.

If one only has a series representation of a function $f(g)$, then the only way to construct the Borel transform is as in Eq.~\eqref{fB}. However, if the function is defined through an action, for example in the $n$-dimensional case
\begin{equation}
    f_S(g) = \int d^n\vec{z}\ e^{-S(\vec{z})/g} \,,
    \label{ffromS}
\end{equation}
then one can compute the Borel transform through a change of variables. Supposing $S(\vec{z})\ge 0$, we have
\begin{equation}
f_S(g) =   \frac{1}{g} \int_0^{\infty} d t\ e^{- \frac{t}{g}} \left[ g \int d^n\vec{z}\ \delta (t - S (\vec{z})) \right]  \,,
\end{equation}
and the Borel transform for $\frac{1}{g} f (g)$ can be read off as
\begin{equation}
  \cB \left[ \frac{1}{g} f_S \right]\!\!(t) 
  =\!\!
  \int d^n \vec{z}\ \delta (t - S (\vec{z})) 
 = \hspace{-1em}\int\limits_{S(\vec{z}) = t} \!\!\!\! d \sigma(\vec{z})\,\frac{1}{|\nabla S(\vec{z})|} 
  \label{Bsum}
\end{equation}
where $d\sigma(\vec{z})$ is the hypersurface measure over the level set $S(\vec{z}) = t$. 
This manipulation identifies the Borel variable $t$ with the action $S$ and shows that saddles of the action, where $\nabla S(\vec{z}_i)=0$, lead to branch points in the Borel plane. 
One can approximate $f_S(g)$ by performing a saddle point approximation around any particular saddle leading to an asymptotic series. Generally, Borel resumming  such a series will not recover the original
function $f_S(g)$. Instead, it will reproduce the integral of $e^{- S(\vec{z})/g}$ over a
different integration contour, namely the Lefschetz thimble passing through the saddle. One can construct a thimble
by moving away from the saddle to regions of asymptotic convergence of the integral. Thimble contours are middle-dimensional in $\mathbb{C}^n$. They can be chosen to have constant imaginary part or can be chosen as any contour in the same relative homology class.
By decomposing the original integration contour $\vec{z} \in \mathbb{R}^n$ into a sum over complex thimbles in $\mathbb{C}^n$, one can then reconstruct the function
through  Borel resummation.
See~\cite{Serone:2017nmd,Tanizaki:2015gpl, Cherman:2014ofa, Dorigoni:2019} for some examples.

\begin{figure}[t!]
    \centering
\includegraphics[width=0.34\textwidth]{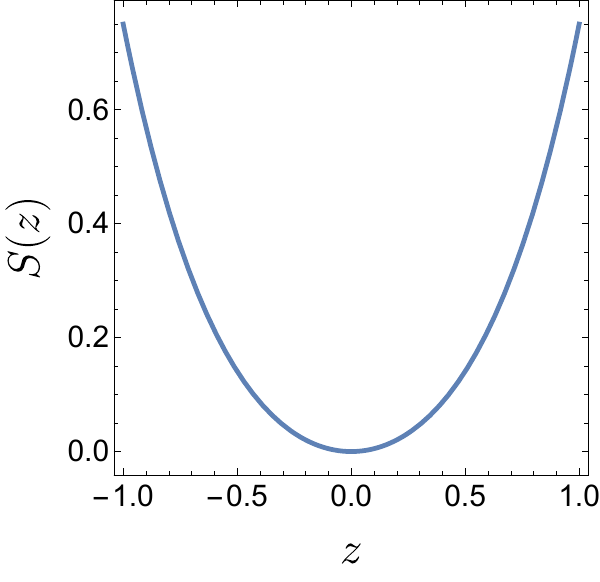}
\includegraphics[width=0.34\textwidth]{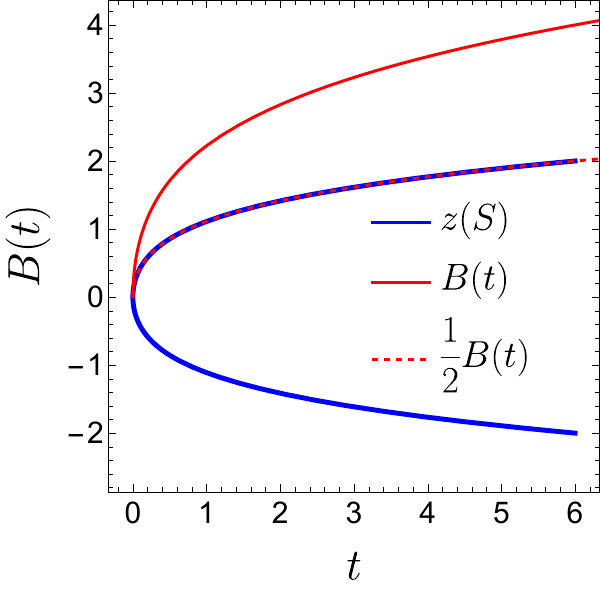}
\\
\includegraphics[width=0.34\textwidth]{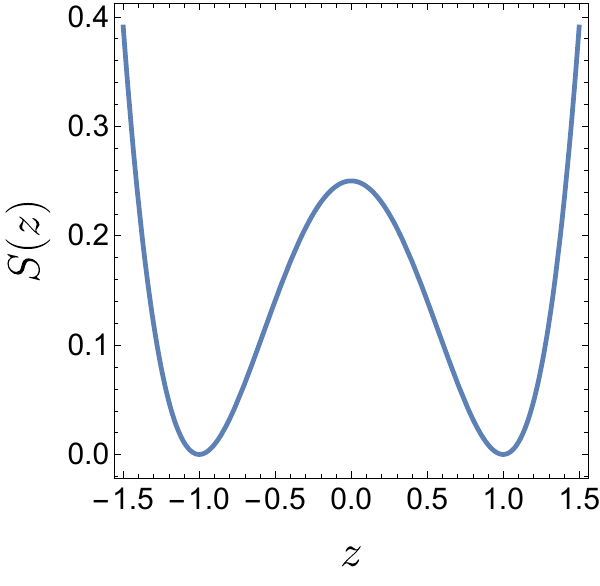}
\includegraphics[width=0.34\textwidth]{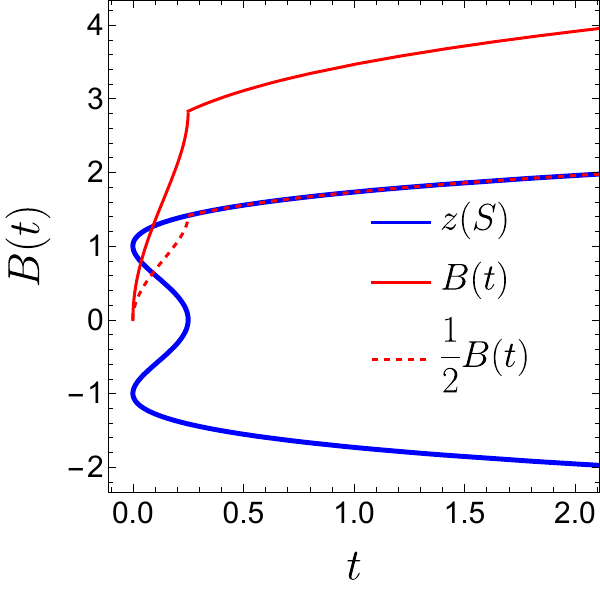}
    \caption{Examples of the correspondence between action and Borel transform. Left shows two quartic actions $S(z) = \frac{1}{2} z^2 +\frac{1}{4} z^4$ and $S(z) = \frac{1}{4}-\frac{1}{2} z^2 +\frac{1}{4}z^4$. Right shows their multivalued inverses $z(S)$ (blue) and Borel transforms (red). The Borel transform is the sum of $|z(S)|$ over different inversion domains.
    }
    \label{fig:stablequartic}
\end{figure}

In one dimension, there is a beautiful duality between the action and the Borel transform.
In 1D,
Eqs.~\eqref{Bsum} and~\eqref{diffB} give
\begin{equation}
  \frac{d}{d t} \cB [f_S] (t) = \sum_{z_i |  S (z_i) = t}
  \left| \frac{1}{S' (z_i)} \right|\,,
\label{E:1d1}
\end{equation}
where the sum is over all points $z_i$ for which $t = S(z_i)$.  The inverse function $z(S) \equiv S^{-1}(z)$ is  multi-valued and can be parsed into single-valued functions on overlapping domains. Substituting $t = S(z)$ into Eq.~\eqref{E:1d1} then gives
\begin{equation}
  \frac{d B (S)}{d S} = \sum_{\text{domains}} \left| \frac{d z(S)}{d S} \right| \,.
\end{equation}
Integrating both sides with respect to $S$ and identifying $S$ with $t$ then shows
\begin{align}
\label{StBz}
B(t) = \sum_{z_i | S(z_i) = t} \pm z_i
\end{align}
where the signs are chosen so that $\pm z_i = \text{sgn}(z_i'(t)) z_i$.  So in 1D, not only is the Borel variable $t$ equal to the action, but the action variable $z$ is equal to the Borel transform 
(with refinements
for  multi-valuedness). From this point of view, the detour into complex coordinates at a Stokes point can be understood as inappropariately continuing one of the domains of $z$ beyond the relevant saddle of $S$.
Some examples of the correspondence between $S(z)$ and $B(t)$ are shown in Fig.~\ref{fig:stablequartic} and discussed in App.~\ref{app:unstable}. 

The higher-dimensional analog of Eq.~\eqref{StBz} follows also from 
Eqs.~\eqref{Bsum} and~\eqref{diffB}
and integration in $t$:
\begin{equation}
    B(t) = \int d^n \vec{z}\ \Theta(t-S(\vec z))\,.
    \label{Bdensity}
\end{equation}
Thus $B(t)$ is the volume of the domain for which $S(\vec{z}) \le t$. A physicist might note that $B'(t)$ is a kind of density of states, describing the volume of paths or fields with action below a given value. A mathematician might identify $B(t)$ 
with the sublevel sets for the Morse function $S : M \to \RR$, where $M$ is the manifold for which $\vec{z}$ are coordinates. Morse theory provides a mapping from critical points of the Morse function to singularities of the Borel transform~\cite{pham1985descente,howls1997hyperasymptotics,Delabaere:2002,witten2011analytic}. Both perspectives are useful.

\subsection{Borel singularities}
Now let us ask: what do singularities in $B(t)$ tell us about $S(\vec{z})$?  Eq.~\eqref{Bdensity} shows us that singularities in $B(t)$ can arise in at least three ways:
\begin{enumerate}[label=(\roman*)]
    \item $S(\vec{z})$ has a critical point at some $\vec{z}_*$ which leads to a branch point in the Borel plane at $t = S(\vec{z}_*)$.
    \item $S(\vec z)$ asymptotes to a constant value $t_*$ as one or more of the coordinates $\vec{z}$ go to infinity (e.g.~saddles at infinity).
    \item  When $t$ hits a certain value $t_*$, there is suddenly an infinite coordinate volume  for which $S(\vec{z}) = t_*$, and as such $B(t)$ is divergent for $t > t_*$.
\end{enumerate}
Instantons in toy 1D models and some QFT instantons like the Fubini instanton in $\phi^4$ theory realize (i). Some examples of this class are discussed in App.~\ref{app:unstable}. We then proceed to discuss case (ii), which is realized by saddles at infinite separation such as the instanton-anti-instanton pairs in the symmetric double-well in QM. Then, we will argue that renormalons realize (iii).
 
Let us now discuss an example of (ii). Recall that in the symmetric double-well potential in QM, or pure Yang-Mills (ML) theory, there are instanton-anti-instanton pairs (henceforth $\ii$) with zero topological charge.  Defining $t_\rI$ be the action of a single instanton, one can find the imprint of the $\ii$ contribution in the Borel transform of $E_0(g)$, the ground state energy of the double-well. Similarly, in QCD, the perturbative series associated with quantities such as the Adler function, is expected to exhibit a Borel pole associated with the $\ii$ pair~\cite{Balitsky:1991sw}. Thus in either case, we expect to encounter quantities whose Borel transform have a singularity at $t=2t_\rI$. Because of Eq.~\eqref{diffB}, whether this singularity is a pole or logarithmic is unimportant for the present discussion. The key feature is that $B(t)$ diverges at $t=2t_\rI$. Letting $z$ be the separation between the instanton and anti-instanton, in either QM or YM the classical action has the form $S(z) = 2t_\rI (1 - e^{-z})$~\cite{Bogomolny:1980ur,Zinn-Justin:1981qzi,balitsky1986collective,Babansky:2000}~\footnote{Note that the functional form of the $\ii$ classical action is only true in the large separation limit. The breaking of scale symmetry, and onset of strong dynamics invalidates the use of this semi-classical method of $\ii$ pairs to describe the physics of pure YM. Here we are only concerned with isolating the asymptotic behavior in perturbative computations stemming from these $\ii$ pairs, without considering any running coupling.}. This reflects the fact that $\ii$ is only a genuine saddle at infinite separation in the full path integral, and is consistent with Eq.~\eqref{StBz} and $B(t)$ blowing up at $t=2t_\rI$. At infinite separation $z$, the action asymptotes to a constant value $S(z) \rightarrow 2t_I$ and the Borel transform of Eq.~(\ref{Bdensity}) is singular. While the full treatment of such a `saddle point at infinity' in the path integral is subtle, we can demonstrate the essential physics with an initial toy 1D integral\footnote{It is really only the large $z$ behavior of Eq.~\eqref{E:f3eq1} which is relevant to the Borel singularity. At large $z$,  Eq.~\eqref{E:f3eq1} arises by integrating out the fluctuations transverse to $z$ in the full path integral and neglecting contributions immaterial to the presence of the Borel singularity.}
\begin{align}
\label{E:f3eq1}
f_{\ii}(g) = \int_0^\infty \!\! dz\, e^{- \frac{2 t_\rI}{g} (1 - e^{-z})}.
\end{align}
Although this integral is divergent, each term in its perturbative expansion around $g=0$ is finite. The Borel transform of the resulting asymptotic series is 
\begin{equation}
B_{\ii}(t) =-\log\left(1-\frac{t}{2t_\rI}\right) \,,    
\end{equation}
which  exhibits the expected singularity at $t = 2 t_\rI$.  

Now let us more properly take account of Eq.~\eqref{E:f3eq1} being divergent.  Inspecting Eq.~\eqref{E:f3eq1}, we notice that the thimble starting at $z=0$ does not end at $z=\infty$ but continues into the complex plane in one of two ways, characteristic of Stokes phenomena~\cite{Behtash:2018voa}.  Deforming $g \to g\pm i\epsilon$ to break the ambiguity and inserting a cutoff at $z = \Lambda$, we can trace a thimble along three contour segments: 
\begin{equation}\nonumber
C_1 =\{0\!<\!z\!<\!\Lambda\}, \quad C_2^\pm = \{ z=\Lambda \pm i y\,|\,0\!<\!y\!<\!\pi\}, \quad C_3^\pm = \{z = \pm i \pi + x\,|\, \Lambda \!>\!x\!>\!-\infty\}.
\end{equation}
In the limit $\Lambda \rightarrow \infty$, the sum of the integral along an entire thimble $C^\pm = C_1 \cup C_2^\pm \cup C_3^\pm$ is
\begin{equation}
    \int_{C^\pm}d z~e^{-\frac{2t_\rI}{g}(1-e^{-z})} =  -e^{-\frac{2t_\rI}{g}} E_1\!\left(-\frac{2t_\rI}{g} \pm i \epsilon\right),
    \label{div1d}
\end{equation}
which is identical to the lateral inverse Borel transform of $B_{\ii}(t)$, as expected. Accordingly, we expect that in full-fledged QCD and QM, the lateral Borel resummation of the leading $\ii$ singularity corresponds to a similarly deformed integration contour into the complex plane or field space.

Finally, we turn to renormalons. Like the $\ii$ instanton, renormalons correspond to singularities in the Borel plane. However, we argue they correspond to mechanism (iii) rather than (ii). Although renormalons require a quantum field theory (to generate the scale anomaly), the mechanism of (iii) can be understood through the lens of an effective action in finite dimensions. Starting from Eq.~\eqref{ffromS}, we can imagine splitting $\vec{z} = (\vec{z}_1, \vec{z}_2)$, e.g.~$\mathbb{R}^n \simeq \mathbb{R}^{n_1} \oplus \mathbb{R}^{n_2}$, and integrating out $\vec{z}_2$.  Then we expect to have a residual integral of the form
\begin{align}\label{E:fseffective}
f_{S}(g) \approx \int d^{n_1}\vec{z}_1\,e^{- \frac{1}{g} \left(S_{0}(\vec{z}_1) + g S_1(\vec{z}_1) + g^2  S_2(\vec{z}_1) + \cdots\right)}\,.
\end{align}
For the moment neglecting the terms at higher than one loop order, the analog of Eq.~\eqref{Bdensity} becomes
\begin{equation}
    B(t) \approx \int d^{n_1} \vec{z_1}\ \Theta(t-S_{0}(\vec z_1))\,e^{- S_1(\vec{z}_1)}\,.
    \label{Bdensityeffective}
\end{equation}
We see that $B(t)$ can diverge for $t$ above some critical $t_*$ if $S_1(\vec{z}_1)$ suddenly becomes unstable (e.g.~not bounded from below) on the hypersurface $S_{0}(\vec{z}_1) = t_*$.  This is an `effective action' manifestation of mechanism (iii). To make the argument sketched above more concrete, we move on to present a quantum field theory example.

\section{Renormalons as saddles}\label{sec:ren_saddle}
To describe how renormalon singularities are generated, we make the link between the usual diagrammatic definition of renormalons and the OPE. For our purposes the renormalon singularities will be probed by looking directly at operator condensates from their path integral definitions.
\subsection{Renormalons in quantum field theories}\label{sec:renormalons_in_qft}

To introduce renormalons in quantum field theories we focus on QCD, though our discussion is applicable to other theories. The classic example of a renormalon is the singularity at $t= \frac{2}{\beta_0}$ in the Borel transform $B_A(t)$ of the Adler function $D(Q^2)$ in 4D Yang-Mills coupled to fermions, where $\beta_0$ is the leading order $\beta$-function coefficient~\cite{tHooft1979,Beneke:1998ui}. The Adler function is defined through the correlation function of two vector currents $j_{\mu}= \bar{q} \gamma_\mu q$ as 
\begin{equation}
    D(Q^2)= 4 \pi^2 \frac{d \Pi(Q^2)}{d Q^2} \quad \text{with} \quad (-i) \int d^4x e^{-iqx} \langle0|T\{j_\mu(x) j_\nu(0)\}|0 \rangle = (q_\mu q_\nu - q^2 g_{\mu \nu}) \Pi(Q^2) \,,
\end{equation}
and the renormalon arises from a chain of vacuum-polarization corrections computed using dimensional-regularization. More generally, in the Adler function renormalons are believed to occur at $t_n=\frac{n}{\beta_0}$ for various integers $n$; while this has been shown for QED, it has not been conclusively shown in QCD \cite{Beneke:1992ch}. When $\beta_0 > 0$ the pole at $t_n$ obstructs the inverse Borel transform and leads to ambiguities. In particular, this imaginary ambiguity is of order $\exp(-t_n/g(Q^2))=(\frac{\Lambda}{Q})^{2n}$, where we used the fact that the one-loop renormalization group equation (RGE) is solved by $g(\mu)^{-1} = 2\beta_0\log(\frac{\mu}{\Lambda})$ . A genuine observable is however naturally unambiguous and therefore the imaginary ambiguities incurred must cancel against appropriate power corrections. 

To see how this cancellation can occur, we start from the the Adler function's OPE
\begin{equation}\label{eq:OPE_Adler}
    D(Q^2) = \sum_{i} C_i\left( \frac{Q^2}{\mu_F^2}\right) \langle \mathcal{O}_i \rangle(\mu_F^2) \left(\frac{1}{Q^2}\right)^{\dimOi/2}\,,
\end{equation}
where $\dimOi$ is the dimension of the operator $\mathcal{O}_i$, $\mu_F$ is a factorization scale, and the sum runs over all operators in the theory that are gauge-invariant Lorentz scalars. This Wilsonian OPE allows one to separate scales. Short-distance contributions (loop momenta $k\gg \mu_F$) are part of the coefficient functions $C_i$, while long-distances (loop momenta $k \ll \mu_F$) only enter the condensates $\langle \mathcal{O}_i\rangle$. If restricted to this rigid cut-off regularization the coefficient functions are free of IR renormalons and the condensates are well-defined\footnote{We can say that the IR renormalon contribution usually present in the coefficient function is now part of the condensate.}. However, when using dimensional regularization, as is the case for most practical calculations, one technically integrates over all scales when calculating the coefficient functions, formally setting $\mu_F \rightarrow 0$. Therefore, in dimensional regularization, the IR renormalons corresponding to asymptotic series in the coupling are present in $C_i$. This subtle interplay of the renormalon appearance in the different versions of the OPE has been understood in significant detail; see Refs. \cite{David:1983gz,David:1985xj,Maiani:1991az,Luke:1994xd,Hoang:2009yr}  To guarantee a well-defined observable, the condensates $\langle \mathcal{O}_i \rangle$ must themselves be ambiguous power corrections. In the example of the Adler function, the IR renormalons computed from diagrammatics in dimensional regularization are all contributions to $C_{1}$, the coefficient function associated with the unit operator. One thus expects operator condensates of dimension $2n$ to cancel the associated ambiguities. Such cancellations remain elusive in QCD, but have been rigorously demonstrated in 2D solvable models \cite{Schubring:2021hrw, Marino:2024uco,Marino:2025ido}.

In the following we will probe the renormalon appearance by focusing on the condensates themselves in scale-invariant theories. We will start from their path integral definition and show how the renormalons materialize as saddle points. Furthermore, we will show that by integrating over the thimble
passing through that saddle point we reproduce the expected form of the imaginary ambiguity. We stress, however, that our aim in this paper is only to reproduce the renormalon, not to diagnose the expected ambiguity cancellation.

\subsection{A renormalon saddle point in the path integral}
In the following we will look at condensates starting from their path integral definition. For instance, the  OPE of the Adler function given in Eq.~(\ref{eq:OPE_Adler}) contains gluonic condensates, with the lowest dimensional operator being $\cO_{GG}(x) = \alpha_s\text{tr}\,[G_{\mu\nu}(x)]^2$, given by 
\begin{equation}
    \langle \cO_{GG}(0) \rangle = \frac{1}{Z} \int [d A_\mu]\, e^{-\frac{1}{\alpha_s}S_0[A_\mu]} \cO_{GG}[A_\mu(0)]  \,.
\end{equation}
To make our exposition general, our starting point is to consider a classically scale-invariant, relativistic  field theory in $D$ Euclidean  dimensions.  Let the field degrees of freedom be denoted by $\phi(x)$ where we suppress possible indices (e.g.~for spin-1 gauge fields) and the classical action by $S_0[\phi(x)]$ after its coupling $g$ is scaled out. Our goal is to compute the expectation value of some operator $\mathcal{O}[\phi(x)] = \mathcal{O}(x)$ with scaling dimension $\Delta_{\mathcal{O}}$, specifically we will compute the condensate
\begin{align}\label{opedef}
\langle \mathcal{O}(0) \rangle = \frac{1}{Z} \int [d\phi]\, e^{-\frac{1}{g}S_0[\phi]} \mathcal{O}[\phi(0)]  \,.
\end{align}
If $\phi$ has classical scaling dimension $\Delta_\phi$, then the action is invariant under the simultaneous dilatations and translations $\phi(x) \to R^{-\Delta_\phi}\phi(\frac{x-x_0}{R})$. In a classically scale-invariant theory, any nontrivial saddle is degenerate, complicating the method of steepest descent. When scale invariance is broken by quantum effects, the degeneracy is lifted by integrating out UV fluctuations to produce the 1-loop effective action.  To see this, 
it will be prudent to use coordinates on field space which reflect these invariances.  In particular, we consider ``$R$-coordinates'' $(R, x_0, \Phi(x))$ which parameterize the space of fields as $\phi(x) =  R^{-\Delta_\phi} \Phi(\frac{x-x_0}{R})$.

A useful basis for $\Phi(x) = \sum_{n=1}^\infty a_n \Phi_n(x)$ is constructed in~\cite{Andreassen:2017rzq,Bhattacharya:2024chz} for 4D scalar fields, which generalizes to higher spin fields.  We can integrate out the UV modes in the decomposition, corresponding to $\Phi_n(x)$ with $n > N$.  Defining $\chi(x) \equiv \sum_{n = 1}^N a_n \Phi_n(x)$, Eq.~\eqref{opedef} becomes 
\begin{align}
\label{E:Oexpect1}
\langle \mathcal{O}(0) \rangle = & \frac{1}{Z} \int [d\chi] \int\! d^{D} x_0 \!\int_{R_0}^\infty \!\!\frac{dR}{R^{D+1}}\,\frac{\mathcal{J}[\chi]}{N_{\text{int}}[\chi]}\,\mathcal{F}_{\mathcal{O}}[\chi] \\
& 
\times \,e^{-\left(\frac{1}{g(\mu)} - 2 \beta_0 \log(\mu R)\right) S_0[\chi] + S_{1}[\chi]}\, \mathcal{O}\left[\frac{1}{R^{\Delta_\phi}}\chi\left(\frac{x_0}{R}\right)\right]\,. \nonumber
\end{align}

Here we implement the saddle point expansion in the renormalized coupling $g(\mu)$, which is the same coupling that would appear in perturbative Feynman diagrams. The renormalized coupling appears at an arbitrary scale $\mu$ such that $g(\mu) \ll 1$, permitting a saddle point approximation to the operator matrix element. Above, the dependence on $\log(\mu R)$ is fixed by RG invariance; $S_1[\chi]$ is the remainder of the effective action; $\mathcal{F}_\cO[\chi]$ comes from the contributions of $\mathcal{O}$ after integrating out the UV modes; and ${\mathcal J}[\chi]$ and $N_{\text{int}}[\chi]$ are respectively parts of the Jacobian and intersection number~\cite{Bhattacharya:2024chz} arising from the transition to 
$R$-coordinates.
The $\beta_0 \log (\mu R)$ that multiplies the classical action is important and universal~\cite{Babansky:2000}.
Depending on the sign of $\beta_0$, the above integral may be UV or IR divergent. To isolate the IR renormalon, we have put a lower cutoff $R_0$ on the $R$ integral. If the integral is convergent at large $R$ then the IR renormalon is still present, but Borel resummable. 

While the integration over $R$ extends into regimes where strong coupling dynamics arise, we emphasize that we intend to merely examine and reproduce the asymptotic nature of perturbative expansions, for instance those associated with the diagrammatic evaluation of Wilson coefficients.  A typical example is the contribution to $C_1$ in the Adler function, the coefficient function associated with the unit operator (introduced in Sec.~\ref{sec:renormalons_in_qft}) whose perturbative evaluation schematically leads to the following
\begin{align}
    C_{\text{pert}} = \sum_{n=0}^{\infty}\int_{0}^{\infty} \frac{dk^2}{k^2}~g(\mu)^{n+1} \ln\left(\frac{\mu^2}{k^2}\right)^{n} F(k^2) \,.
\end{align}
  In practice, dimensional regularization lets the integration over $k$ (the equivalent of $R^{-1}$) go into regions where $g(k)$ ceases to be perturbative, but nonetheless the asymptotic series associated with $C_{\text{pert}}$ (in $g(\mu)$) remains purely perturbative in nature. Thus, in what follows we are examining the effects or presence of the renormalon in the perturbative evaluation of physical quantities in the path integral, and do not comment on their role (or lack thereof) in the true non-perturbative dynamics of the theory. 

We next  integrate over  $x_0$  and extract the $R$ dependence of the operator $\mathcal{O}$ by dimensional analysis.  Changing coordinates to $\rho = \log(\mu R)$, the integral in Eq.~\eqref{E:Oexpect1} then becomes
\begin{align}
\label{Oexpect2}
 \mu^{\dimO}  \!  \! \int [d\chi]  \!\int_{ \log(\mu R_0)}^\infty\!\! d\rho\,\mathcal{F}[\chi]\,e^{-\frac{1}{g(\mu)}S_0[\chi] -\rho ( \Delta_{\mathcal{O}} - 2 \beta_0 S_0[\chi])}\,,
\end{align}
where 
\begin{equation}
  \mathcal{F}[\chi] \equiv \frac{1}{Z}\frac{\mathcal{J}[\chi]}{N_{\text{int}}[\chi]} \mathcal{F}_{\mathcal{O}}[\chi] \,e^{- S_1[\chi]} \,.  
\end{equation}
Crucially, the effective action in Eq.~\eqref{Oexpect2} has saddle points where
\begin{align}\label{E:saddlepts}
\big(1 - 2 \beta_0 g(\mu)\rho\big) \frac{\delta S_0}{\delta \chi} = 0\,,\qquad S_0[\chi] = \frac{\Delta_{\mathcal{O}}}{2\beta_0}\,,
\end{align}
which can be solved by
\begin{equation}
    \rho = \frac{1}{2 \beta_0 g(\mu)} \quad   \text{or equivalently} \quad R= \frac{1}{\mu}\exp\left(\frac{1}{2\beta_0 g(\mu)}\right) = \frac{1}{\Lambda} \,.
\end{equation}
This is the renormalon. It has the expected classical action which, as desired, depends on the operator scaling dimension \footnote{
In~\cite{Babansky:2000}, Babansky and Balitsky used constrained instanton methods to show that field configurations  along a valley between the vacuum saddle and $\ii$ encode the renormalon singularity in the Borel plane. Although we were inspired by their approach, we note that only field configurations near the vacuum are actually needed for their calculation; the existence of a valley and instantons are irrelevant. Our more general perspective allows for the identification of the renormalon saddle and the thimbles in field space associated with lateral Borel resummation.} and an appropriate scale. In particular, the action-Borel correspondence implies that the saddle point action $S_0 = \dimO /2\beta_0$ is linked to poles in the Borel plane at $t= \dimO /2\beta_0$. In QCD the gluonic operators have even dimensions, $\dimO=2n$, such that the associated poles are located at $t_n = n/\beta_0$. Their location matches the poles in the Borel transform of the Adler function.

Having located the renormalon as a saddle point of the 1-loop effective action, it is natural to ask what happens beyond one-loop. Normally, a renormalon is {\it defined} as the asymptotic growth of a perturbative series computed with bubble-chains in the 1-loop approximation. There can be other sources of growth from other graphs, but the renormalon is defined through the bubble chains. Similarly, when writing $\Lambda = \mu \exp(-\frac{1}{2\beta_0 g(\mu)})$, this relation holds only at one-loop, or equivalently, is a definition of $\Lambda$ based on 1-loop running. Thus, while one could attempt to expand our analysis to higher order, there is no reason to expect the higher order results to preserve the nature of the renormalon. Nevertheless, with
two-loop running, we find that the renormalon action $S_0[\chi]$ remains the same, although the nature of the singularity changes. This is consistent with 
subleading effects found with a diagrammatic approach~\cite{Grunberg:1995vx,Peris:1996in}. Our path integral approach therefore provides an alternative path towards higher-order corrections but there are no clear expectations for what aspects of the renormalon will or should survive such an analysis.

\subsection{Thimbles and the renormalon ambiguity}
Having identified the renormalon saddle point, it is now natural to ask how it contributes to the full path integral of Eq.~(\ref{opedef}). To answer this question, it is essential to first refine which integration path we should be considering. First, we note that the integral over $\rho$ in Eq.~\eqref{Oexpect2} is divergent when the classical action surpasses the renormalon action: $S_0[\chi] >\frac{\Delta_{\mathcal{O}}}{2\beta_0}$. And, therefore, to make a sensible integral one must deform the integration contour onto a thimble (or chain of thimbles). 
Performing a change of variables in Eq.~\eqref{Oexpect2} from $S_0[\chi]$ to $s$ and provisionally deforming the $(\rho,s)$ part of the integration contour onto a new contour $C$, we have
\begin{align}
\label{E:Borel1}
\mu^{\dimO}  \! \! \int_C ds \wedge d\rho\, e^{-\frac{s}{g(\mu)} - \rho (\dimO - 2\beta_0 s)} \widetilde{B}(s)
\end{align}
where we defined
\begin{align}
\label{E:Btilde1}
\widetilde{B}(s) = 
\int\limits_{S_0[\chi] = s} \! [d \sigma(\chi)] \,\frac{\mathcal{F}[\chi]}{\big|\frac{\delta S_0[\chi]}{\delta \chi}\big|}
\end{align}
and $[d\sigma(\chi)]$ is the measure over the level set $S_0[\chi]=s$.  To describe which contours $C$ are relevant in Eq.~(\ref{E:Borel1}), it is sufficient to consider a simpler toy 2D integral\footnote{This integral is a special case of Eq.~(\ref{E:fseffective}) with $S_0(s,\rho) = s$ and $S_1(s,\rho) = (t_R-s)\rho$.}:
\begin{equation}
   f_R(g)= \int_0^\infty \! d s \int_0^\infty \! d\rho \,e^{-\frac{1}{g} s - (t_R - s)\rho } \,.
   \label{E:fR1}
\end{equation} Indeed this 2D integral represents Eq.~(\ref{E:Borel1}) before the deformation to the new contour $C$ and where we set $\tilde{B}(s)=1$ for simplicity.
This integral is divergent but we will refine it shortly.  Formally we can compute the Borel transform of $f_R(g)$ as
\begin{align}
\label{E:BR1}
B_R(t) = \int_0^\infty \! ds  \int_0^\infty \! d\rho\,\Theta(t - s)\,e^{-(t_R - s) \rho}\,,
\end{align}
as in Eq.~\eqref{Bdensityeffective}.  For $t < t_R$ the Borel transform is $B_R(t) = - \log(1 - \frac{t}{t_R})$ while if $t \geq t_R$ then $B_R(t)$ is divergent, since the exponential integral in $\rho$ destabilizes. The action in Eq.~\eqref{E:fR1} has a non-trivial ``renormalon'' saddle at $(s,\rho)=(t_R, 1/g)$ on the boundary of the instability, illustrating mechanism (iii) by which the singularities can arrise in the Borel transform.

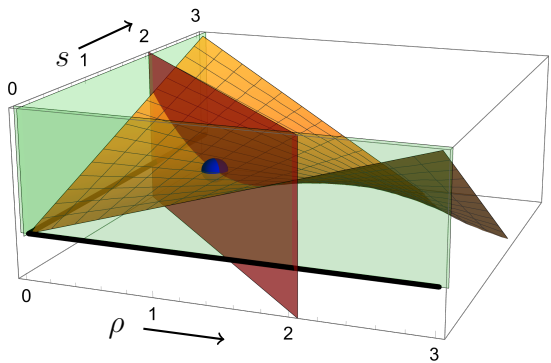
\begin{figure}[t!]
    \centering
  \begin{tikzpicture}
        \node[anchor=south west,inner sep=0] (image) at (0,0) {\includegraphics[width=0.5\textwidth]{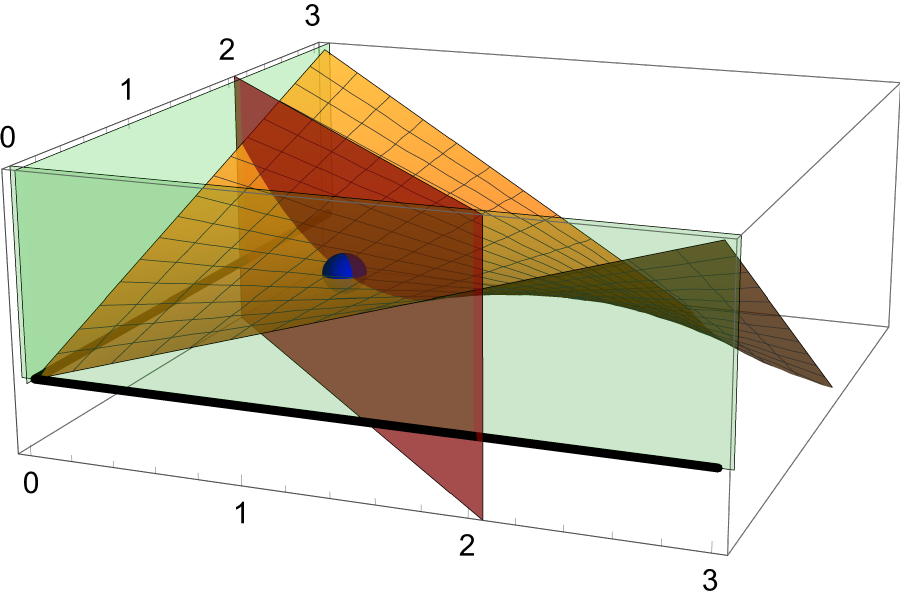}};
        \begin{scope}[x={(image.south east)},y={(image.north west)}]

            \node at (0.2, 0.1) {\large $\rho$};
            \draw[->, thick] (0.25, 0.1) -- (0.4, 0.07);
            \node at (0.1, 0.85) {\large $s$};
            \draw[->, thick] (0.13, 0.88) -- (0.23, 0.95);
        \end{scope}
    \end{tikzpicture}
    \caption{Real part of the action $S_{\text{tot}}=s+g(t_R-s)\rho$ shown in orange for $g=t_R=1$. This action has a non-trivial `renormalon' saddle at $s=\rho=1$ (blue dot). The planes represent thimbles where the vertical axis is reinterpreted as the direction  $w=\im~s = \im~\rho$. The steepest ascent surface from the boundary (black lines) is real until it merges with the red plane by the saddle point. This boundary thimble can be deformed into the green half-planes for ease of integration. The renormalon thimble is shown in red.
    }
    \label{fig:int2d}
\end{figure}

The divergence of the Borel transform for $t \ge t_R$ and of $f_R$ in Eq.~\eqref{E:fR1} is similar to the divergence of the $\ii$ instanton integral in QCD. Near the origin $(s,\rho)= (0,0)$, the total action $S_{\text{tot}} = s + g(t_R - s) \rho$ grows in the positive $s$ and $\rho$ directions. Near the renormalon the total action grows in one real direction, where $\re\, s =- \re\, \rho$, and one imaginary, where $\im\, s = \im\, \rho$.
Thus the thimble which begins on the original boundary $\Omega = (s=0, \rho\in \RR^+)\cup(s\in \RR^+, \rho=0)$ must move into a complex direction to avoid the Stokes point at the saddle. As such, we should really consider a contour in the appropriate relative homology class, such as one which simply moves up from $\Omega$ in the imaginary direction as shown in Fig.~\ref{fig:int2d}. Writing $s=x+i w$ and $\rho =y+i w$, this ``boundary'' thimble is over the $(x,y,w)$ surfaces  $C_1 = (\RR^+,0,\RR^+)$ and $C_2 = (0, \RR^+,\RR^+)$, and letting $C = C_1 \cup C_2$ we have the refined integral
\begin{equation}
   \int_{C} ds\wedge d\rho\,e^{-\frac{1}{g}S_{\text{tot}}} = -e^{ - \frac{t_R}{g}} E_1 \!\left( - \frac{t_R}{g} + i \epsilon \right).\label{E1form}
\end{equation}
Integrating along this 2D thimble which avoids the renormalon Stokes point reproduces the lateral Borel resummation of the $f_R(g)$ asymptotic series.  The right-hand side has a similar form as an instanton-anti-instanton pair, but arises by mechanism (iii) rather than mechanism (ii). Aside from the boundary thimble, we observe that one can also integrate along a thimble $C_R$ passing through the renormalon saddle point. In $(x,y,w)$ coordinates this {\it $R$-thimble} is the surface where $-\infty < x - t_R = 1/g -y <\infty$
 and $-\infty < w< \infty$. Integrating along this contour gives twice the imaginary part of Eq.~\eqref{E1form}.

Returning to the integral of Eq.~(\ref{E:Borel1}), we now see that we have two thimble contours to consider. If $C$ is the boundary thimble, containing the region where where $R = R_0$ and $s=\chi=0$, the result will be analogous to Eq.~(\ref{E1form}) and correspond to the lateral Borel resummation of the renormalon asymptotic series, with imaginary part 
proportional to $i \mu^{\dimO} \exp[-\Delta_{\mathcal{O}}/(2\beta_0 g(\mu))]$.\footnote{Although $\chi=0$ is not a solution to Eq.~\eqref{E:saddlepts}, the free theory is a saddle point on the $R=R_0$ hypersurface. Such boundary saddles can be studied with Picard-Lefschetz theory just as saddles in unbounded spaces~\cite{Delabaere:2002}.}The sign of the imaginary part is linked to the choice of contour deformation.  We note in passing that if the theory has instantons as well as renormalons, these will appear as branch points in $\widetilde{B}(s)$.
If $C$ is the $R$-thimble, which passes through the renormalon, the imaginary part of the integral should also be proportional to the same expression. This imaginary part has precisely the form that one would expect from the condensates. Indeed, using the one-loop definition for $g(\mu)$, we have  $i \mu^{\dimO} \exp[-\Delta_{\mathcal{O}}/(2\beta_0 g(\mu))] = i \Lambda^{\dimO}$, which matches the form of the typical renormalon ambiguities obtained from diagrammatics.

For a concrete example of a renormalon in a physical theory, consider the trace of the gluon field strength tensor squared in QCD: $\cO_{GG}(x) = \alpha_s\text{tr}\,[G_{\mu\nu}(x)]^2$. This is the leading operator in the OPE for the Adler function. The integral of its expectation value along the $R$ thimble gives
\begin{equation} 
\label{eq:R_thimble_val}
\langle \cO_{GG}(0)\rangle_{R} \approx \frac{i \pi}{\beta_0} \mu^4 \,e^{-\frac{2}{\beta_0 g(\mu)}} \widetilde{B}^* =  \frac{i \pi}{\beta_0} \Lambda^4 \widetilde{B}^*
\end{equation}
where
\begin{equation}
   \widetilde{B}^*=
   \frac{1}{2\pi}\int dz \,d\bar{z}\,e^{- z \bar{z}} \widetilde{B}(z) 
\end{equation}
with $z= s - \frac{2}{\beta_0}, \bar{z}= \frac{1}{2\beta_0 g}-\rho$, such that the integral is taken over the red plane depicted in Fig.~\ref{fig:int2d}.\footnote{To all orders in the expansion of $\widetilde{B}(z)$ around $z=0$ we have $\widetilde{B}^* = \widetilde{B}(t_R)$.} 
Eq.~\eqref{eq:R_thimble_val} is the expected structure of the ambiguity associated with the leading IR renormalon at $t=2/\beta_0$ of the Adler function~\cite{Mueller:1984vh,Zakharov:1992bx,Neubert:1994vb}, and $\widetilde{B}^*$ is non-vanishing (barring cancellations from different orders in perturbation theory, which we do not expect).  Indeed, all the directions around the renormalon are either collective coordinates which are understood or directions with positive eigenvalues or a single direction with a negative eigenvalue. Notice that the right-hand side of Eq.~\eqref{eq:R_thimble_val} is pure imaginary; while this was previously observed in perturbative analyses, our analysis clarifies that the $i$ is due to the $R$-thimble being a complex integration contour.
In addition, we note that since our analysis is done at the path integral level, it does not require any large $N$ or large $N_f$ limits as are usually common when isolating renormalon bubble-chain diagrams. Therefore, our identification of the renormalon in a non-Abelian theory provides an alternate probe into its existence beyond na\"ive non-Abelianization from Feynman diagrams~\cite{Broadhurst:1994se,Beneke:1994qe}.

\section{Conclusion} \label{sec:conclusions}
In this paper we have approached the study of renormalons using the language of path integrals, which is typically reserved for instantons. Both instantons and renormalons correspond to singularities in the Borel transform of asymptotic series, but they differ in the nature of those singularities. This distinction is best understood by considering series expansions of functions defined as exponential integrals, where there exists a duality between the action and the Borel transform of that series. Indeed, one can extract the Borel transform solely from knowledge of the action, and vice versa, one can reconstruct an action whose exponential integral reproduces a given asymptotic series. From that duality, we concluded that singularities in the Borel transform are tied to various behaviors of the associated action. In particular, we argued that instanton singularities are typically associated with the action possessing a critical point or asymptoting to a constant value along a given coordinate. In contrast, renormalons correspond to an action where the coordinate volume becomes infinite at a given point. Notably, we showed that in classically scale-invariant theories this coordinate is identified as the collective coordinate for scale invariance. Looking at the path integral corresponding to operator condensates, we further illustrated how renormalons can also be understood as saddle points -- not of the classical action, but of the effective action.

Identifying renormalons as saddles in the path integral provides a framework for examining their use in precision perturbative computations in QCD via the OPE. It remains to perform precision checks of the putative correspondence between renormalon ambiguities and the true non-perturbative energy scale in confining theories. Admittedly, answering this question in QCD-like theories is exceptionally challenging, and much work remains to be done. Nonetheless, having a path integral understanding of renormalons as saddles opens a broad new avenue to examine their role and origin in quantum field theory.

\acknowledgments
We would like to thank Marco Serone, Semon Rezchikov and Frank Wilczek for valuable conversations. This work was supported in part by the U.S. Department of Energy under contract DE-SC0013607.

\appendix

\section{Borel-Action examples \label{app:unstable}}
\noindent 
This appendix gives some examples of how one can reconstruct a 1D action $S(z)$ from the Borel transform $B(t)$ using Eq.~\eqref{StBz}.

For a first example, consider the stable quartic potential $S_1(z) =
\frac{1}{2} z^2 + \frac{1}{4} z^4$ as shown in Fig.~\ref{fig:stablequartic}. Then
\begin{equation}
  f_{S_1} (g) = \int_{- \infty}^{\infty} d z\ e^{- \frac{1}{2 g} z^2 - \frac{1}{4
  g} z^4} = \frac{1}{\sqrt{2}} \,e^{\frac{1}{8 g}} \,\mathcal{K}_{\frac{1}{4}}
  \!\left( \frac{1}{8 g} \right) \,.
\end{equation}
This function is non-analytic. Its expansion approaching $g = 0$ from the
positive real direction is asymptotic:
\begin{equation}
  f_1 (g) \sim 
  \sum_{n = 0}^{\infty}
  \sqrt{2g} (- g)^n \frac{\Gamma \left( \frac{1}{2} + 2 n
  \right)}{n!} \,.
\end{equation}
Its Borel transform is 
\begin{equation}
  B_1 (t) = \sum_{n = 0}^{\infty} \frac{\Gamma ( \frac{1}{2} + 2 n)}{\Gamma \left( n + \frac{3}{2}
  \right)n!} \sqrt{2 t} (- t)^n  = 2 \sqrt{\sqrt{1 + 4 t} - 1} \,. \label{Bgf}
\end{equation}
This Borel transform  has a branch point at $t = - \frac{1}{4}$. It is Borel summable and the inverse Borel transform exactly
reproduces $f_{S_1} (g)$. 

Alternatively, we could compute the Borel transform
using Eq.~\eqref{StBz}. Solving $S (z) = t$ for $z$ gives two real roots $z_\pm
= \pm \sqrt{\sqrt{1 + 4 t} - 1}$. Summing these two roots agrees with Eq.~\eqref{Bgf}, corroborating Eq.~\eqref{StBz}. 
Conversely, we can construct the action from the Borel transform by solving Eq.~\eqref{StBz} for $S$ and $z$. Expecting at least two domains for a stable action, we set $B (t) = 2 z$ which leads to $S = t = \frac{z^2}{2} + \frac{z^4}{4}$ and the original action is reproduced.

Next we consider the unstable quartic, with action $S_2(z) = \frac{1}{4} - \frac{1}{2} z^2 + \frac{1}{4} z^4$. Then 
\begin{align}
  f_{S_2} (g) 
  &= 
 \int_{- \infty}^{\infty} d z\ e^{- \frac{1}{g} \left[ \frac{1}{4} -\frac{1}{2} z^2 + \frac{1}{4} z^4 \right]}\\
&
  = \frac{\pi}{2} \,e^{- \frac{1}{8
  g}} \left[ \mathcal{I}_{- \frac{1}{4}}\! \left( \frac{1}{8 g} \right) +
  \mathcal{I}_{\frac{1}{4}} \!\left( \frac{1}{8 g} \right) \right] \,.
\end{align}
For this action, $S_2(z) = t$ has two real solutions $z = \pm \sqrt{1 + 2
\sqrt{t}} $ for any positive $t$ and when $0 < t < \frac{1}{4}$ two additional real
solutions $z = \pm \sqrt{1 - 2 \sqrt{t}}$. The Borel transform is
then immediately computed using Eq.~\eqref{StBz}:
\begin{equation}
  B_2 (t) 
  = \underbrace{- 2 \sqrt{1 - 2 \sqrt{t}}\ \theta \left(
  \frac{1}{4} - t \right)}_{\text{domain in the middle}}
  + \underbrace{2 \sqrt{1 + 2 \sqrt{t}}}_{\text{domain on the ends}} \,.
  \label{B2domains}
\end{equation}
This action and its Borel transform are also shown in
Fig.~\ref{fig:stablequartic} where one can see their relationship graphically.

The unstable quartic has three saddle points at $z=-1,0,1$. Performing the saddle point expansion around each saddle point gives an asymptotic series whose Borel resummation corresponds to the integral along the Lefschetz thimble passing through the saddle. For example, with $z = 1+x$, we compute
\begin{align}
  f_2^{(1)} (g) &= \int_{- \infty}^{\infty} d x\  e^{- \frac{x^2}{g}} \sum_n
  \frac{1}{n!} \left( - \frac{x^3 + \frac{1}{4} x^4}{g} \right)^n \\
  &= 
  \sum_{n = 0}^{\infty} g^{n+\frac{1}{2}} \frac{\Gamma \left( \frac{1}{2} + 2 n \right)}{n!} 
\end{align}
with Borel transform
\begin{equation}
  \cB[f_2^{(1)}] = \sum_{n = 0}^{\infty} t^{n + \frac{1}{2}} \frac{\Gamma \left(
  \frac{1}{2} + 2 n \right)}{n! \,\Gamma \left( \frac{3}{2} + n \right)} =
  \sqrt{2 - 2 \sqrt{1 - 4 t \pm i \epsilon}} \,.
  \label{Bf21}
\end{equation}
The branch point at $t=\frac{1}{4}$ corresponds to the action of the Stokes point at $z=0$, which requires a choice of $\pm i$ in the deformation of the thimble contour. The Borel transform of the expansion around the $z=-1$ saddle also gives Eq.~\eqref{Bf21}. 

For the $z=0$ saddle, we can pull out the $e^{-S(0)/g} \ne 1$ factor and compute from the series
\begin{align}
  \cB \left[ e^{\frac{1}{4 g}} f_2^{(0)} \right] &= \pm i   \sum_{n
  = 0}^{\infty} \sqrt{2t} (- t)^{n}
  \frac{\Gamma \left( \frac{1}{2} + 2 n \right)}{\Gamma \left( n + \frac{3}{2} \right)n!} \\
  &= \pm 2 i \sqrt{\sqrt{1 + 4
  t} - 1} \,.
\end{align}
Technically, the expansion of $f_{S_2}(g)$ around the $z=0$ saddle is a trans-series where the powers of $g$ are supplemented by $e^{-1/g}$ terms. These can be compensated for by shifting the $t$ integration domain so that
\begin{equation}
 \cB [f_2^{(0)}] (t) = 2 \sqrt{1 - 2 \sqrt{t} \pm i \epsilon}\,  \theta \left( t - \frac{1}{4} \right) \,.
\end{equation}
We then check that  $2\cB [f_2^{(1)}]+\cB [f_2^{(0)}]$ agrees with Eq.~\eqref{B2domains}. Moreover, we also check that the inverse Borel transform of each separate Borel transform reproduces the integral along the associated Lefschetz thimble.

\bibliographystyle{JHEP}
\bibliography{Refs.bib}

\providecommand{\href}[2]{#2}\begingroup\raggedright\begin{thebibliography}{10}

\bibitem{Lipatov:1976ny}
L.N.~Lipatov, \emph{{Divergence of the Perturbation Theory Series and the
  Quasiclassical Theory}}, {\emph{Sov. Phys. JETP} {\bfseries 45} (1977) 216}.

\bibitem{tHooft1979}
G.~'t~Hooft in \emph{The Whys of Subnuclear Physics}, A.~Zichichi, ed.,
  (Boston, MA), pp.~943--982, Springer US (1979),
  \href{https://doi.org/10.1007/978-1-4684-0991-8_17}{DOI}.

\bibitem{Lautrup:1977hs}
B.E.~Lautrup, \emph{{On High Order Estimates in QED}},
  \href{https://doi.org/10.1016/0370-2693(77)90145-9}{\emph{Phys. Lett. B}
  {\bfseries 69} (1977) 109}.

\bibitem{Gross:1974jv}
D.J.~Gross and A.~Neveu, \emph{{Dynamical Symmetry Breaking in Asymptotically
  Free Field Theories}},
  \href{https://doi.org/10.1103/PhysRevD.10.3235}{\emph{Phys. Rev. D}
  {\bfseries 10} (1974) 3235}.

\bibitem{Novikov:1984ac}
V.A.~Novikov, M.A.~Shifman, A.I.~Vainshtein and V.I.~Zakharov,
  \emph{{Two-Dimensional Sigma Models: Modeling Nonperturbative Effects of
  Quantum Chromodynamics}},
  \href{https://doi.org/10.1016/0370-1573(84)90021-8}{\emph{Phys. Rept.}
  {\bfseries 116} (1984) 103}.

\bibitem{Beneke:1993yn}
M.~Beneke and V.I.~Zakharov, \emph{{The First infrared renormalon in QED}},
  \href{https://doi.org/10.1016/0370-2693(93)91090-A}{\emph{Phys. Lett. B}
  {\bfseries 312} (1993) 340}.

\bibitem{Marino:2019fvu}
M.~Mari\~no and T.~Reis, \emph{{A new renormalon in two dimensions}},
  \href{https://doi.org/10.1007/JHEP07(2020)216}{\emph{JHEP} {\bfseries 07}
  (2020) 216} [\href{https://arxiv.org/abs/1912.06228}{{\ttfamily
  1912.06228}}].

\bibitem{Marino:2019eym}
M.~Mari\~no and T.~Reis, \emph{{Renormalons in integrable field theories}},
  \href{https://doi.org/10.1007/JHEP04(2020)160}{\emph{JHEP} {\bfseries 04}
  (2020) 160} [\href{https://arxiv.org/abs/1909.12134}{{\ttfamily
  1909.12134}}].

\bibitem{Schubring:2021hrw}
D.~Schubring, C.-H.~Sheu and M.~Shifman, \emph{{Treating divergent perturbation
  theory: Lessons from exactly solvable 2D models at large N}},
  \href{https://doi.org/10.1103/PhysRevD.104.085016}{\emph{Phys. Rev. D}
  {\bfseries 104} (2021) 085016}
  [\href{https://arxiv.org/abs/2107.11017}{{\ttfamily 2107.11017}}].

\bibitem{Shifman2015}
M.~Shifman, \emph{Resurgence, operator product expansion, and remarks on
  renormalons in supersymmetric yang-mills theory},
  \href{https://doi.org/10.1134/S1063776115030115}{\emph{Journal of
  Experimental and Theoretical Physics} {\bfseries 120} (2015) 386}.

\bibitem{Shifman:2022xsa}
M.~Shifman, \emph{{Infrared renormalons in supersymmetric theories}},
  \href{https://doi.org/10.1103/PhysRevD.107.045002}{\emph{Phys. Rev. D}
  {\bfseries 107} (2023) 045002}
  [\href{https://arxiv.org/abs/2211.05090}{{\ttfamily 2211.05090}}].

\bibitem{Argyres:2012vv}
P.~Argyres and M.~Unsal, \emph{{A semiclassical realization of infrared
  renormalons}},
  \href{https://doi.org/10.1103/PhysRevLett.109.121601}{\emph{Phys. Rev. Lett.}
  {\bfseries 109} (2012) 121601}
  [\href{https://arxiv.org/abs/1204.1661}{{\ttfamily 1204.1661}}].

\bibitem{Argyres:2012ka}
P.C.~Argyres and M.~Unsal, \emph{{The semi-classical expansion and resurgence
  in gauge theories: new perturbative, instanton, bion, and renormalon
  effects}}, \href{https://doi.org/10.1007/JHEP08(2012)063}{\emph{JHEP}
  {\bfseries 08} (2012) 063} [\href{https://arxiv.org/abs/1206.1890}{{\ttfamily
  1206.1890}}].

\bibitem{Dunne:2012zk}
G.V.~Dunne and M.~\"Unsal, \emph{{Continuity and Resurgence: towards a
  continuum definition of the $\mathbb{CP}$(N-1) model}},
  \href{https://doi.org/10.1103/PhysRevD.87.025015}{\emph{Phys. Rev. D}
  {\bfseries 87} (2013) 025015}
  [\href{https://arxiv.org/abs/1210.3646}{{\ttfamily 1210.3646}}].

\bibitem{Dunne:2012ae}
G.V.~Dunne and M.~Unsal, \emph{{Resurgence and Trans-series in Quantum Field
  Theory: The CP(N-1) Model}},
  \href{https://doi.org/10.1007/JHEP11(2012)170}{\emph{JHEP} {\bfseries 11}
  (2012) 170} [\href{https://arxiv.org/abs/1210.2423}{{\ttfamily 1210.2423}}].

\bibitem{Maiani:1991az}
L.~Maiani, G.~Martinelli and C.T.~Sachrajda, \emph{{Nonperturbative
  subtractions in the heavy quark effective field theory}},
  \href{https://doi.org/10.1016/0550-3213(92)90528-J}{\emph{Nucl. Phys. B}
  {\bfseries 368} (1992) 281}.

\bibitem{Beneke:1998ui}
M.~Beneke, \emph{{Renormalons}},
  \href{https://doi.org/10.1016/S0370-1573(98)00130-6}{\emph{Phys. Rept.}
  {\bfseries 317} (1999) 1}
  [\href{https://arxiv.org/abs/hep-ph/9807443}{{\ttfamily hep-ph/9807443}}].

\bibitem{grozin1997higher}
A.~Grozin and M.~Neubert, \emph{Higher-order estimates of the chromomagnetic
  moment of a heavy quark}, {\emph{Nuclear Physics B} {\bfseries 508} (1997)
  311}.

\bibitem{Hoang:2009yr}
A.H.~Hoang, A.~Jain, I.~Scimemi and I.W.~Stewart, \emph{{R-evolution: Improving
  perturbative QCD}},
  \href{https://doi.org/10.1103/PhysRevD.82.011501}{\emph{Phys. Rev. D}
  {\bfseries 82} (2010) 011501}
  [\href{https://arxiv.org/abs/0908.3189}{{\ttfamily 0908.3189}}].

\bibitem{Luke:1994xd}
M.E.~Luke, A.V.~Manohar and M.J.~Savage, \emph{{Renormalons in effective field
  theories}}, \href{https://doi.org/10.1103/PhysRevD.51.4924}{\emph{Phys. Rev.
  D} {\bfseries 51} (1995) 4924}
  [\href{https://arxiv.org/abs/hep-ph/9407407}{{\ttfamily hep-ph/9407407}}].

\bibitem{Beneke:2016cbu}
M.~Beneke, P.~Marquard, P.~Nason and M.~Steinhauser, \emph{{On the ultimate
  uncertainty of the top quark pole mass}},
  \href{https://doi.org/10.1016/j.physletb.2017.10.054}{\emph{Phys. Lett. B}
  {\bfseries 775} (2017) 63}
  [\href{https://arxiv.org/abs/1605.03609}{{\ttfamily 1605.03609}}].

\bibitem{Hoang:2017suc}
A.H.~Hoang, A.~Jain, C.~Lepenik, V.~Mateu, M.~Preisser, I.~Scimemi et~al.,
  \emph{{The MSR mass and the $
  \mathcal{O}\left({\Lambda}_{\mathrm{QCD}}\right) $ renormalon sum rule}},
  \href{https://doi.org/10.1007/JHEP04(2018)003}{\emph{JHEP} {\bfseries 04}
  (2018) 003} [\href{https://arxiv.org/abs/1704.01580}{{\ttfamily
  1704.01580}}].

\bibitem{Babansky:2000}
A.~Babansky and I.~Balitsky, \emph{Renormalons as dilatation modes in the
  functional space},
  \href{https://doi.org/10.1103/PhysRevLett.85.4211}{\emph{Phys. Rev. Lett.}
  {\bfseries 85} (2000) 4211}.

\bibitem{Serone:2017nmd}
M.~Serone, G.~Spada and G.~Villadoro, \emph{{The Power of Perturbation
  Theory}}, \href{https://doi.org/10.1007/JHEP05(2017)056}{\emph{JHEP}
  {\bfseries 05} (2017) 056}
  [\href{https://arxiv.org/abs/1702.04148}{{\ttfamily 1702.04148}}].

\bibitem{Tanizaki:2015gpl}
Y.~Tanizaki, \emph{{Study on sign problem via Lefschetz-thimble path
  integral}}, .

\bibitem{Cherman:2014ofa}
A.~Cherman, D.~Dorigoni and M.~Unsal, \emph{{Decoding perturbation theory using
  resurgence: Stokes phenomena, new saddle points and Lefschetz thimbles}},
  \href{https://doi.org/10.1007/JHEP10(2015)056}{\emph{JHEP} {\bfseries 10}
  (2015) 056} [\href{https://arxiv.org/abs/1403.1277}{{\ttfamily 1403.1277}}].

\bibitem{Dorigoni:2019}
D.~{Dorigoni}, \emph{{An introduction to resurgence, trans-series and alien
  calculus}}, \href{https://doi.org/10.1016/j.aop.2019.167914}{\emph{Annals of
  Physics} {\bfseries 409} (2019) 167914}
  [\href{https://arxiv.org/abs/1411.3585}{{\ttfamily 1411.3585}}].

\bibitem{pham1985descente}
F.~Pham, \emph{La descente des cols par les onglets de lefschetz, avec vues sur
  gauss-manin}, {\emph{Ast{\'e}risque} {\bfseries 130} (1985) 11}.

\bibitem{howls1997hyperasymptotics}
C.~Howls, \emph{Hyperasymptotics for multidimensional integrals, exact
  remainder terms and the global connection problem}, {\emph{Proceedings of the
  Royal Society of London. Series A: Mathematical, Physical and Engineering
  Sciences} {\bfseries 453} (1997) 2271}.

\bibitem{Delabaere:2002}
E.~Delabaere and C.J.~Howls, \emph{{Global asymptotics for multiple integrals
  with boundaries}},
  \href{https://doi.org/10.1215/S0012-9074-02-11221-6}{\emph{Duke Mathematical
  Journal} {\bfseries 112} (2002) 199 }.

\bibitem{witten2011analytic}
E.~Witten, \emph{Analytic continuation of chern-simons theory}, {\emph{AMS/IP
  Stud. Adv. Math} {\bfseries 50} (2011) 347}.

\bibitem{Balitsky:1991sw}
I.I.~Balitsky, \emph{{Instanton induced asymptotics of perturbative series for
  R (e+ e- ---{\ensuremath{>}} hadrons) and Lambda (QCD)}},
  \href{https://doi.org/10.1016/0370-2693(91)91685-O}{\emph{Phys. Lett. B}
  {\bfseries 273} (1991) 282}.

\bibitem{Bogomolny:1980ur}
E.B.~Bogomolny, \emph{{CALCULATION OF INSTANTON - ANTI-INSTANTON CONTRIBUTIONS
  IN QUANTUM MECHANICS}},
  \href{https://doi.org/10.1016/0370-2693(80)91014-X}{\emph{Phys. Lett. B}
  {\bfseries 91} (1980) 431}.

\bibitem{Zinn-Justin:1981qzi}
J.~Zinn-Justin, \emph{{Multi - Instanton Contributions in Quantum Mechanics}},
  \href{https://doi.org/10.1016/0550-3213(81)90197-8}{\emph{Nucl. Phys. B}
  {\bfseries 192} (1981) 125}.

\bibitem{balitsky1986collective}
I.~Balitsky and A.V.~Yung, \emph{Collective-coordinate method for quasizero
  modes}, {\emph{Physics Letters B} {\bfseries 168} (1986) 113}.

\bibitem{Behtash:2018voa}
A.~Behtash, G.V.~Dunne, T.~Schaefer, T.~Sulejmanpasic and M.~\"Unsal,
  \emph{{Critical Points at Infinity, Non-Gaussian Saddles, and Bions}},
  \href{https://doi.org/10.1007/JHEP06(2018)068}{\emph{JHEP} {\bfseries 06}
  (2018) 068} [\href{https://arxiv.org/abs/1803.11533}{{\ttfamily
  1803.11533}}].

\bibitem{Beneke:1992ch}
M.~Beneke, \emph{{Large order perturbation theory for a physical quantity}},
  \href{https://doi.org/10.1016/0550-3213(93)90554-3}{\emph{Nucl. Phys. B}
  {\bfseries 405} (1993) 424}.

\bibitem{David:1983gz}
F.~David, \emph{{On the Ambiguity of Composite Operators, IR Renormalons and
  the Status of the Operator Product Expansion}},
  \href{https://doi.org/10.1016/0550-3213(84)90235-9}{\emph{Nucl. Phys. B}
  {\bfseries 234} (1984) 237}.

\bibitem{David:1985xj}
F.~David, \emph{{The Operator Product Expansion and Renormalons: A Comment}},
  \href{https://doi.org/10.1016/0550-3213(86)90279-8}{\emph{Nucl. Phys. B}
  {\bfseries 263} (1986) 637}.

\bibitem{Marino:2024uco}
M.~Marino and R.~Miravitllas, \emph{{Trans-series from condensates}},
  \href{https://arxiv.org/abs/2402.19356}{{\ttfamily 2402.19356}}.

\bibitem{Marino:2025ido}
M.~Marino, \emph{{Anatomy of the simplest renormalon}},
  \href{https://arxiv.org/abs/2504.12044}{{\ttfamily 2504.12044}}.

\bibitem{Andreassen:2017rzq}
A.~Andreassen, W.~Frost and M.D.~Schwartz, \emph{{Scale Invariant Instantons
  and the Complete Lifetime of the Standard Model}},
  \href{https://doi.org/10.1103/PhysRevD.97.056006}{\emph{Phys. Rev. D}
  {\bfseries 97} (2018) 056006}
  [\href{https://arxiv.org/abs/1707.08124}{{\ttfamily 1707.08124}}].

\bibitem{Bhattacharya:2024chz}
A.~Bhattacharya, J.~Cotler, A.~Dersy and M.D.~Schwartz, \emph{{Collective
  coordinate fix in the path integral}},
  \href{https://doi.org/10.1103/PhysRevD.110.116023}{\emph{Phys. Rev. D}
  {\bfseries 110} (2024) 116023}
  [\href{https://arxiv.org/abs/2402.18633}{{\ttfamily 2402.18633}}].

\bibitem{Grunberg:1995vx}
G.~Grunberg, \emph{{Renormalons and fixed points}},
  \href{https://doi.org/10.1016/0370-2693(96)00061-5}{\emph{Phys. Lett. B}
  {\bfseries 372} (1996) 121}
  [\href{https://arxiv.org/abs/hep-ph/9512203}{{\ttfamily hep-ph/9512203}}].

\bibitem{Peris:1996in}
S.~Peris and E.~de~Rafael, \emph{{On renormalons and Landau poles in gauge
  field theories}},
  \href{https://doi.org/10.1016/0370-2693(96)01053-2}{\emph{Phys. Lett. B}
  {\bfseries 387} (1996) 603}
  [\href{https://arxiv.org/abs/hep-ph/9603359}{{\ttfamily hep-ph/9603359}}].

\bibitem{Mueller:1984vh}
A.H.~Mueller, \emph{{On the Structure of Infrared Renormalons in Physical
  Processes at High-Energies}},
  \href{https://doi.org/10.1016/0550-3213(85)90485-7}{\emph{Nucl. Phys. B}
  {\bfseries 250} (1985) 327}.

\bibitem{Zakharov:1992bx}
V.I.~Zakharov, \emph{{QCD perturbative expansions in large orders}},
  \href{https://doi.org/10.1016/0550-3213(92)90054-F}{\emph{Nucl. Phys. B}
  {\bfseries 385} (1992) 452}.

\bibitem{Neubert:1994vb}
M.~Neubert, \emph{{Scale setting in QCD and the momentum flow in Feynman
  diagrams}}, \href{https://doi.org/10.1103/PhysRevD.51.5924}{\emph{Phys. Rev.
  D} {\bfseries 51} (1995) 5924}
  [\href{https://arxiv.org/abs/hep-ph/9412265}{{\ttfamily hep-ph/9412265}}].

\bibitem{Broadhurst:1994se}
D.J.~Broadhurst and A.G.~Grozin, \emph{{Matching QCD and HQET heavy - light
  currents at two loops and beyond}},
  \href{https://doi.org/10.1103/PhysRevD.52.4082}{\emph{Phys. Rev. D}
  {\bfseries 52} (1995) 4082}
  [\href{https://arxiv.org/abs/hep-ph/9410240}{{\ttfamily hep-ph/9410240}}].

\bibitem{Beneke:1994qe}
M.~Beneke and V.M.~Braun, \emph{{Naive nonAbelianization and resummation of
  fermion bubble chains}},
  \href{https://doi.org/10.1016/0370-2693(95)00184-M}{\emph{Phys. Lett. B}
  {\bfseries 348} (1995) 513}
  [\href{https://arxiv.org/abs/hep-ph/9411229}{{\ttfamily hep-ph/9411229}}].

\end{thebibliography}\endgroup

\end{document}